# Phase-based Edge Coupling: Modeling Overlapping Functional Architectures and High-Order Edge-to-Edge Interactions in EEG Networks

*Chengcheng Wu, Zichen Shi, Jiayang Xu, Shanbao Tong, Senior Member, IEEE, Xiaoli Guo, Senior Member, IEEE*

***Abstract*—Network neuroscience has traditionally relied on node-centric approaches, which are inherently limited in capturing the brain's overlapping functional architectures and higher-order edge-to-edge interactions. The emerging edge-centric paradigm addresses this gap by shifting the analytical unit from nodes to edges. However, while edge-centric models have been established for DTI and fMRI, comparable EEG methodologies have yet to be developed. In this study, we proposed a Phase-based Edge-centric Coupling (PEC) algorithm for EEG network modelling, where edge time series are derived from instantaneous phase differences of EEG signals and edge-to-edge coupling is quantified via normalized inner product. Simulation tests demonstrated the robustness of PEC algorithm, achieving high accuracy (<10% relative error) in coupling estimation and discriminative capacity (AUROC > 0.7) against spurious coupling. When provided sufficient data length (≥16 cycles), PEC outperformed traditional Phase Node-centric Coupling (PNC) in both individual identifiability and cognitive-state identifiability, enabling superior characterization of stable individual neural signatures and dynamic task-dependent functional configurations. Furthermore, unlike PNC's hard-partitioned modularity, PEC captured highly overlapping yet functionally specialized modular structures and connector hubs, aligning with task-engaged cognitive processes. Applied to dyadic interactions, PEC identified distinct cross-brain modules encoding symmetric and asymmetric neural coordination. These findings establish PEC as a valid and powerful tool for EEG network modelling, providing new insights into the brain's complex, multifunctional organization at both intra-brain and inter-brain networks.**

## I. Introduction

THE brain is a quintessential complex system, whose remarkable computational power arises from intricate interactions among billions of neural elements. Network neuroscience has provided a powerful mathematical language to describe this complexity, modeling the brain as a graph where nodes represent neural units and edges denote their statistical or physical dependencies [1]. This framework has been instrumental in uncovering fundamental principles of brain organization, such as modular segregation and small-world efficiency, which balance specialized local processing with integrated global communication [2, 3]. However, the dominant analytical paradigm in this field, the node-centric connectivity analysis, confines analytical units to vertices and thus inherently limits our ability to fully capture the brain's dynamic and multifaceted functional architecture.

One of the fundamental limitations of the node-centric paradigm is its rigid modularity modeling. Standard node-centric community detection algorithms assign each node to a single, non-overlapping module, producing a "hard-partitioned" representation of brain organization [4]. However, a growing body of evidence demonstrated that the brain's functional modules are highly overlapping [5-7]. Many cortical regions, such as the precuneus and anterior insula, are not dedicated to a single cognitive process but serve as flexible hubs involved in multiple functional systems—from the default mode network to the frontoparietal control network—depending the tasks [6]. These flexible hubs, also known as connector hubs, facilitate inter-community connections to promote information integration across different functional modules [5]. This multifunctionality is not a system bug but an evolutionary adaptation, enabling efficient allocation of finite

This work was supported by the National Natural Science Foundation of China (No. 62371285) and Fundamental Research Funds for the Central Universities (No. YG2023QNB21). (Corresponding authors: Xiaoli Guo)

Chengcheng Wu, Zichen Shi, Jiayang Xu, Shanbao Tong, Xiaoli Guo are with School of Biomedical Engineering, Shanghai Jiao Tong University, Shanghai, China. (e-mail: augustus@sjtu.edu.cn; shizichen@sjtu.edu.cn; xjy990610@sjtu.edu.cn; stong@sjtu.edu.cn; meagle@sjtu.edu.cn)

This work involved human subjects in the research. Approval of all ethical and experimental procedures and protocols was granted by the Review Board of Ethics Committee of Shanghai Children's Medical Center (SCMCIRB-K2020108-1 and SCMCIRB-K2023099-1) and the Ethics Committee of Scientific Research Ethics Committee of the School of Medicine, Hangzhou Normal University (20221111001).

anatomical resources. However, by enforcing exclusive community membership, node-centric methods systematically obscure this critical feature of brain organization, leading to simplified and potentially misleading functional mappings. Although overlapping community detection algorithms, such as Clique Percolation Method (CPM) [8], the Stochastic Variational Inference for the Mixed-membership Stochastic Block Model (SVINET) [9], the Affiliation Graph Model (AGMFIT) [10], have been proposed, their efficacy remains fundamentally constrained by the node-centric framework, making them difficult to achieve balance between excessive community overlap (blurring partition distinctiveness) and high proportions of unclassified vertices (undermining coverage), resulting in unstable partition results [11].

A more fundamental limitation is that the node-centric paradigm is intrinsically incapable of modeling higher-order edge-to-edge interactions, which are crucial for understanding the complex functions of the brain. Traditional node-centric network metrics (e.g., degree, clustering coefficient) essentially describe the topological properties of vertices, while overlooking connections as fundamental units of information processing. However, emerging evidence suggested that brain functional organization is not solely encoded in region-to-region connectivity, but fundamentally relies on the coordinated activity of connection ensembles serving integrated computations [12]. The functional characteristics of vertices in brain networks are largely determined by their connectivity relationships, which inherently contain rich structural and dynamic information [12]. By aggregating edges to form multi-level collective structures such as motifs, fingerprints, and edge communities, cross-system collaborative patterns can be revealed [12-14]. Furthermore, analysis of edge time series indicated that transient patterns of functional connectivity in the brain are driven by synchronous fluctuations of grouped edges rather than simple summation of vertex activities [12]. Therefore, in-depth exploration of brain function requires not only a vertex-centric perspective but also an edge-oriented approach.

Meanwhile, edge connectivity analysis is not only applicable to intra-brain networks but also holds significant relevance for inter-brain interaction research. In traditional node-centric paradigm, we can only examine the connectivity relationships between vertices belonging to different brains. However, inter-brain network research indicated that brain functional organization is reflected not only in inter-regional connectivity patterns but also in the synergy, coupling, and higher-order network structures formed through cross-brain connections [15, 16]. Furthermore, edge connectivity analysis enables the examination of functional relationships between “intra-brain edges” and “inter-brain edges”, providing novel analytical approaches for hyper-network research spanning single-brain and multi-brain systems.

To model overlapping functional architectures and high-order edge-to-edge interactions, an edge-centric framework has been introduced to complex network theory [17]. Unlike node-centric approaches, edge-centric connectivity analysis treats "edges" (i.e., connections between vertices) as fundamental units and functional entities with intrinsic temporal dynamics. By directly quantifying the coupling between edges, for instance, through their time-varying co-fluctuations, researchers can construct an “edge network” to analyze network organizational structures and functional characteristics. Crucially, this framework inherently reveals overlapping modular structures, i.e., clustering edges into non-overlapping “link communities” and projecting them onto constituent nodes automatically generates overlapping node communities, thereby reflecting the brain’s multifunctional architecture [17]. Furthermore, this perspective provides a universal toolkit for probing higher-order network dynamics, as it directly addresses the question of how connection ensembles collectively support cognitive processes through their coordinated activities.

This new framework has significantly advanced our understanding of the human connectome. In structural imaging, edge-centric analysis of diffusion tensor imaging (DTI) data has revealed that white matter pathways constitute a highly interwoven, overlapping scaffold underpinning regional multifunctionality [18]. In functional magnetic resonance imaging (fMRI), an edge-centric functional connectivity (eFC) algorithm has been developed to capture co-fluctuations between brain-region pairs by tracking the dynamics of connection ensembles, demonstrating superior subject-specific fingerprinting and overlapping modular modeling compared to traditional nodal approaches [4, 19]. These methods have undergone sustained advancement through both model refinement and clinical application. Novelli et al. systematically elucidated the theoretical foundations of edge-centric functional connectivity from a mathematical modeling perspective, and proposed a unified framework for formalizing the generation, measurement, and statistical inference of edge time series, thereby providing crucial support for the standardization and reproducibility of edge-centric methods [20]. Jones et al. integrated edge time series with generalized linear models (GLMs), establishing an "Edge-Based GLM" analytical framework that enables real-time tracking of higher-order cognitive states at (sub)-second time scales, significantly enhancing the analytical capacity of functional connectivity for fine-grained cognitive processes [21]. Zamani Esfahlani et al. applied edge-centric dynamic network analysis to clinical neuroscience, revealing that children with Autism Spectrum Disorder exhibit abnormal dynamic stability in specific edge communities during social cognitive tasks [22].

However, as a novel analytic framework, the current approaches of the edge-centric analysis remain predominantly confined to DTI and fMRI modalities. Electroencephalography (EEG), with its millisecond temporal resolution, offers the ideal modality for capturing the rapid, real-time co-fluctuations between edges. The absence of an EEG-based edge-centric approach thus constitutes a critical methodological gap in this area.

In this study, we developed a Phase-based Edge-centric Coupling (PEC) algorithm—a novel pipeline for EEG edge connectivity analysis, with comprehensive validation and tests.

(1) The accuracy, sensitivity and specificity of PEC in detecting phase-synchronized edge pairs were tested using simulated EEG data with ground-truth edge coupling. (2) The performances of PEC in individual identifiability and cognitive-state identifiability were evaluated using differential identifiability, given that superior functional connectivity algorithms must capture both stable individual-specific neural signatures and dynamic task-dependent functional configurations [23]. (3) The performance of PEC for community detection was tested to demonstrate its advantage in capturing overlapping modules that align with the brain's multifunctional architecture. (4) Finally, the PEC algorithm was applied to a dual-brain EEG dataset to characterize the edge coupling between two brain networks.

## II. Algorithm Design

### A. Phase-based Node-centric Coupling (PNC)

Traditional Phase-based Node-centric Coupling (PNC) approaches in EEG analysis estimate the phase synchronization between narrow-band signals derived from paired nodes (either electrodes or reconstructed brain regions). Among such approaches, the Phase Locking Value (PLV) is a widely adopted metric that quantifies the consistency of instantaneous phase differences between two signals over time [24]. To derive the PLV (Fig. 1a), the EEG signals from each node are first transformed into instantaneous phase time series via the Hilbert transform, and then the PLV between two nodes (e.g., $i$ and $j$) is defined as:

$$PLV_{i,j} = \frac{1}{N}\left|\sum_{t=1}^{N} exp^{i\left(\phi_i(t)-\phi_j(t)\right)}\right|, \quad (1)$$

where $\phi_i(t)$ and $\phi_j(t)$ represent the instantaneous phases of nodes $i$ and $j$ at time point $t$, and $N$ denotes the number of time points. The PLV ranges between 0 and 1, with higher values indicating stronger phase synchronization. A value of 1 occurs only when the phase difference between two nodes remains perfectly constant throughout the time series, while values approaching 0 indicate random phase relationships.

### B. Phase-based Edge-centric Coupling (PEC)

Our Phase-based Edge-centric Coupling (PEC) algorithm extends the theoretical framework of PLV by reorienting the analytical unit from nodes to edges (Fig. 1b). Specifically, for each pair of nodes (e.g., $i$ and $j$), the complex exponential of their instantaneous phase difference at each time point $t$ is used to construct their edge time series $\varepsilon_{<i,j>}(t)$:

$$\varepsilon_{<i,j>}(t) = exp^{i\left(\phi_i(t)-\phi_j(t)\right)}. \quad (2)$$

Notably, this edge time series is identical to the instantaneous component in the PLV, but here serves as the foundational element for edge dynamics. Once the edge time series is obtained, functional connectivity between two edges (e.g., $\varepsilon_{<i,j>}(t)$ and $\varepsilon_{<k,l>}(t)$) is quantified as their consistency of instantaneous phase differences across time, using the normalized inner product (NIP), a complex-domain metric analogous to cosine similarity [25]:

$$NIP\left(\varepsilon_{<i,j>}, \varepsilon_{<k,l>}\right) = \frac{\left|\varepsilon_{<i,j>} \cdot {\varepsilon_{<k,l>}}^*\right|}{\sqrt{\varepsilon_{<i,j>} \cdot {\varepsilon_{<i,j>}}^*} \cdot \sqrt{\varepsilon_{<k,l>} \cdot {\varepsilon_{<k,l>}}^*}} = \frac{1}{N}\left|\sum_{t=1}^{N} exp^{i\left[\left(\phi_i(t)-\phi_j(t)\right)-\left(\phi_k(t)-\phi_l(t)\right)\right]}\right|, \quad (3)$$

where ${\varepsilon_{<i,j>}}^*$ and ${\varepsilon_{<k,l>}}^*$ denote the complex conjugate of $\varepsilon_{<i,j>}$ and $\varepsilon_{<k,l>}$.

While aiming to capture phase synchronization between undirected edges, the NIP must account for bidirectional edge configurations:

$$NIP\left(\varepsilon_{<j,i>}, \varepsilon_{<k,l>}\right) = NIP\left(\varepsilon_{<i,j>}, \varepsilon_{<l,k>}\right) = \frac{1}{N}\left|\sum_{t=1}^{N} exp^{i\left[\left(\phi_i(t)-\phi_j(t)\right)+\left(\phi_k(t)-\phi_l(t)\right)\right]}\right|, \quad (4)$$

$$NIP\left(\varepsilon_{<j,i>}, \varepsilon_{<l,k>}\right) = NIP\left(\varepsilon_{<i,j>}, \varepsilon_{<k,l>}\right) = \frac{1}{N}\left|\sum_{t=1}^{N} exp^{i\left[\left(\phi_i(t)-\phi_j(t)\right)-\left(\phi_k(t)-\phi_l(t)\right)\right]}\right|. \quad (5)$$

To obtain an undirectional coupling metric, we define the PEC as the maximum NIP value across these two configurations:

$$PEC_{(i,j),(k,l)} = max\left(\begin{array}{c}\frac{1}{N}\left|\sum_{t=1}^{N} exp^{i\left[\left(\phi_i(t)-\phi_j(t)\right)-\left(\phi_k(t)-\phi_l(t)\right)\right]}\right|, \\ \frac{1}{N}\left|\sum_{t=1}^{N} exp^{i\left[\left(\phi_i(t)-\phi_j(t)\right)+\left(\phi_k(t)-\phi_l(t)\right)\right]}\right|\end{array}\right). \quad (6)$$

Similar to PLV, PEC also ranges between 0 and 1, with higher values indicating stronger coupling between edges. Code of PEC is available at this repository: https://github.com/JAWu600/Phase-based-Edge-centric-Coupling.

## III. Simulations

### A. EEG Dataset for Simulation

To validate the accuracy, sensitivity and specificity of the PEC algorithm in detecting phase synchronization between edge pairs, we designed two simulation tests based on real EEG signals. The dataset for simulation (**Dataset 1**) includes 1 min eye-open resting-state EEG data from 45 healthy adults (6 males and 39 females, average age = 45.83 ± 14.35 years) recorded via a 32-channel wireless EEG system (NeuSen.W32, Neuracle Technology Co., Ltd., China) at a sampling rate of 1000 Hz. Electrode montage followed the international 10-20 system, with FCz as reference, AFz as ground, and electrode impedances maintained below 10 kΩ. The dataset was collected in our previous study [26] approved by the Ethics Committee of Scientific Research Ethics Committee of the School of Medicine, Hangzhou Normal University (20221111001).

EEG Preprocessing was conducted using MATLAB (The MathWorks, Natick, MA, USA) and the EEGLAB toolbox [27]. First, EEG signals were band-pass filtered at 1-30 Hz using a FIR filter, followed by a 50 Hz notch filter to remove the power line interference. Then, noisy electrodes and highly contaminated segments were removed, and artifacts such as eye movements, muscle artifacts, and generic discontinuities were removed using independent component analysis (ICA).

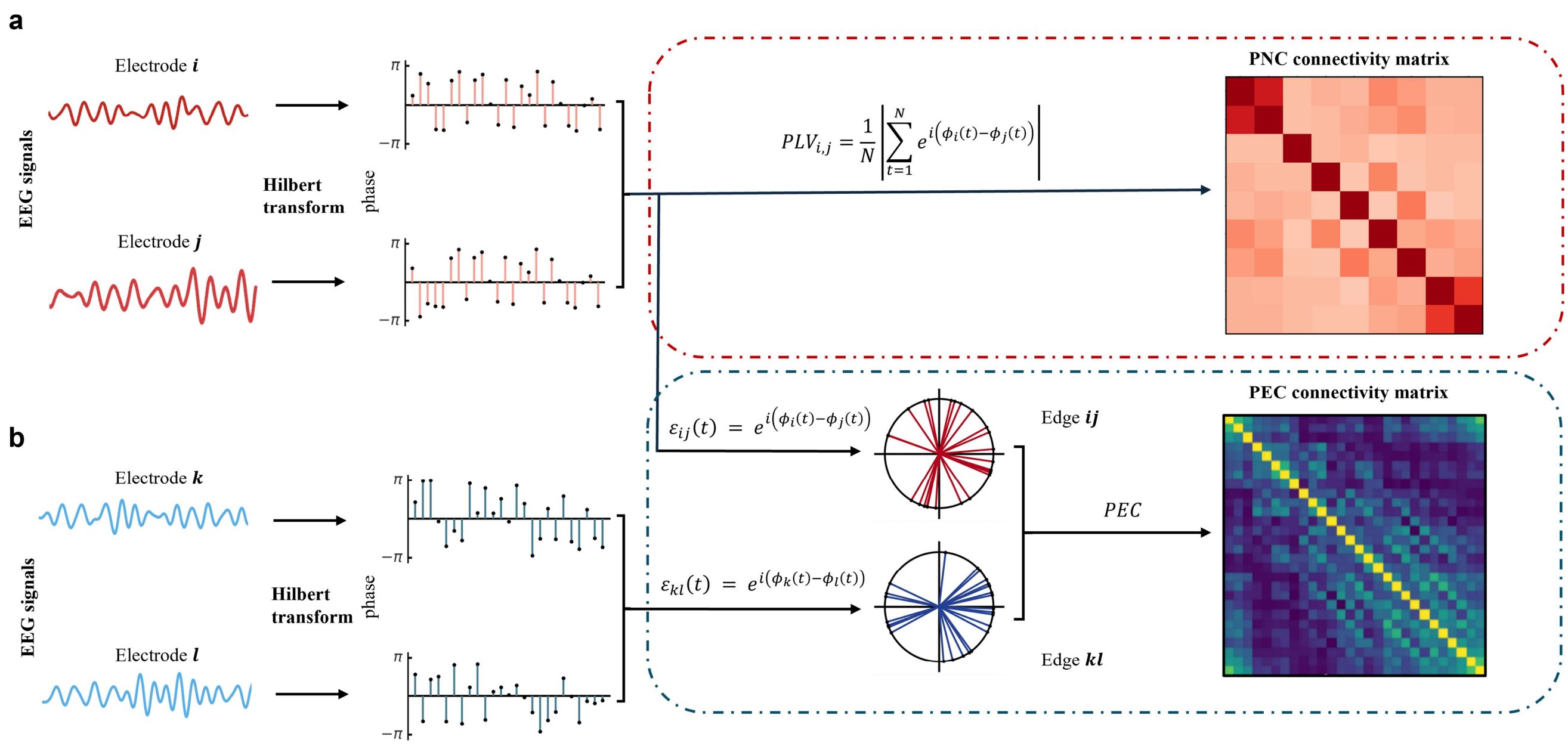

Fig. 1. Derivation of PNC (PLV) and PEC algorithms. (a) The PNC algorithm (PLV) first obtains the phase time series from the original narrow-band signals through the Hilbert transform, and then calculates the PLV from the phase time series of the two vertices to quantify their consistency of instantaneous phase differences across time. (b) Our PEC algorithm first constructs the edge time series of the corresponding edges $\varepsilon_{<i,j>}(t)$ and $\varepsilon_{<k,l>}(t)$ using the complex exponential of instantaneous phase difference of node pairs at each time point $t$, and then calculates the PEC of the two edges based on the normalized inner product (NIP) of the corresponding edge time series. Similar to PNC, PEC quantifies the consistency of instantaneous phase differences between edges across time.

## B. Generation of Simulation Data

Firstly, using the EEG signals in **Dataset 1**, we generated edge pairs (each consisting of a reference edge and a corresponding target edge) with predefined synchronization coefficients through controlled phase manipulation. We randomly selected 2 EEG channels (e.g., $i$ and $j$) of one subject from the dataset and extracted their narrowband-filtered signals in the frequency band of interest. Applying the Hilbert transform, we yielded instantaneous phase time series $\phi_i(t)$ and $\phi_j(t)$ for the reference edge (between nodes $i$ and $j$). Wethen generated two instantaneous phase time series $\phi_k(t)$ and $\phi_l(t)$ for the target edge (between nodes $k$ and $l$) by introducing controlled phase noise:

$$\phi_k(t) = \phi_i(t) + \frac{\eta(t)}{2}, \quad (7)$$

$$\phi_l(t) = \phi_j(t) - \frac{\eta(t)}{2}, \quad (8)$$

where $\eta(t)$ represents phase noise drawn from the von Mises distribution ~ Von Mises($\mu$,$\kappa$), a circular probability distribution suited for phase data for its compatibility with the periodic nature of neural oscillations [28]. Its probability density function is given by:

$$f(x) = \frac{exp^{\kappa cos(x-\mu)}}{2\pi I_0(\kappa)}, \quad (9)$$

where $\mu$ is the directional parameter, $\kappa$ is the concentration parameter, and $I_0(\kappa)$ is the modified Bessel function of the first kind of order zero [29]. The reference and target edges were then defined as:

$$\varepsilon_{<i,j>}(t) = exp^{\mathrm{i}(\phi_i(t)-\phi_j(t))}, \quad (10)$$

$$\varepsilon_{<k,l>}(t) = exp^{\mathrm{i}(\phi_k(t)-\phi_l(t))} = exp^{\mathrm{i}(\phi_i(t)-\phi_j(t)+\eta(t))}. \quad (11)$$

Critically, the phase synchronization between the reference edge and target edge was precisely controlled by the concentration parameter $\kappa$ of the phase noise distribution. We defined the ground-truth synchronization coefficient $\lambda$ as the expected value of the PEC of the edge pairs, which can be analytically derived as [28]:

$$\lambda = \frac{I_1(\kappa)}{I_0(\kappa)}, \quad (12)$$

where $I_1(\kappa)$ is the modified Bessel function of the first kind of order one. This coefficient $\lambda$ ranges from 0 (completely desynchronized) to 1 (perfectly synchronized).

## C. Simulation 1: Accuracy Assessment

This test aimed to validate the accuracy of PEC in estimating different levels of synchronization coefficients. We varied $\lambda$ from 0 to 1 with increments of 0.1 (using 0.02 and 0.99 instead of 0 and 1), solving for the corresponding $\kappa$ values. For each $\lambda$, we generated 100 pairs of edge time series using noise samples from Von Mises($\mu$, $\kappa$), computed their PEC values, and calculated the relative error between the estimated synchronization coefficient and the theoretical ground-truth $\lambda$.

Gaussian white noise was superimposed onto the EEG signals, and the relative errors of the PEC algorithm was evaluated across frequency bands (delta [1-4 Hz], theta [4-8 Hz], alpha [8-12 Hz], and beta [12-25 Hz]) under varying signal lengths (500 ms, 1 s, 2 s, 4 s, and 8 s), coupling coefficients ($\lambda$ = 0.02, 0.1, 0.2, 0.3, 0.4, 0.5, 0.6, 0.7, 0.8, 0.9, 0.99), and signal-to-noise (SNR) conditions (2 dB, 5 dB, 10 dB, 20 dB, and ∞, Fig. 2). It can be noted that PEC underestimated the couplings (estimates < ground-truth) with the influence of noises (Fig. 2a). Quantitative relative error analysis revealed that while estimation errors generally increased with EEG frequency and decreased SNR, robust performance was consistently observed (Fig. 2b). Specifically, in the low-frequency delta band, the relative errors remained <10% across all conditions except at the extremely weak coupling condition of $\lambda = 0.02$, and typically fell <5% when SNR ≥ 5. Theta band errors were generally <12% (excluding $\lambda = 0.02$), and dropped <8% at SNR ≥ 5 and <5% at SNR ≥ 10. In the alpha band, the relative errors remained <10% except at $\lambda = 0.02$ and the low-SNR conditions of SNR = 2, while SNR ≥ 10 yielded errors <5%. In the beta band, the relative errors remained <10% except at $\lambda$ = 0.02 and SNR = 2 or 5, with SNR ≥ 20 achieving errors <5%. These results confirmed high accuracy of the PEC algorithm in estimating edge coupling strength, though the performance degraded under extremely weak coupling and low-SNR conditions.

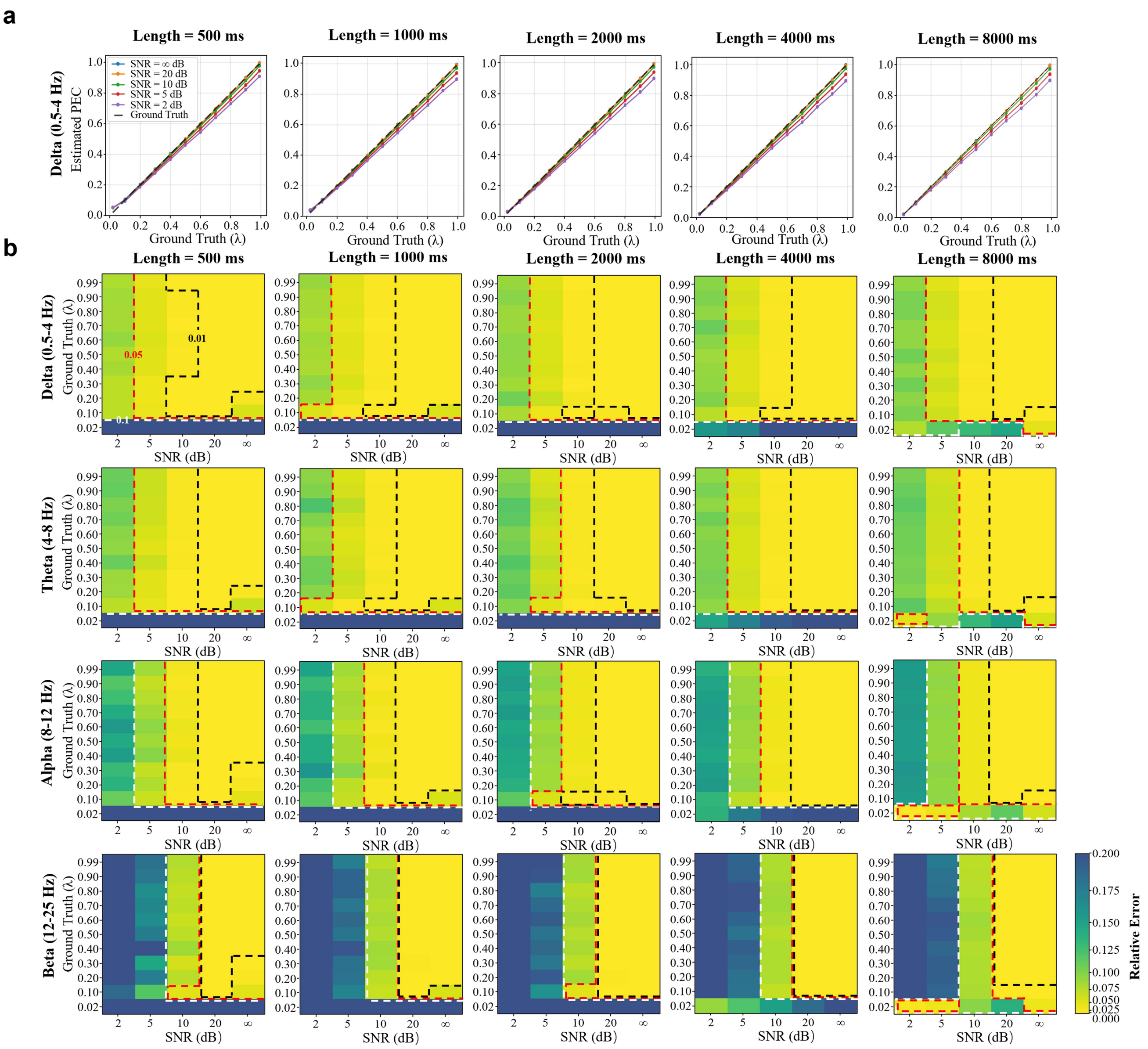

Fig. 2. The PEC estimates and their relative errors against simulated ground-truth coupling coefficients across delta, theta, alpha, and beta bands under varying signal lengths (500 ms, 1000 ms, 2000 ms, 4000 ms, 8000 ms), ground-truth coupling coefficients ($\lambda$ = 0.02, 0.1, 0.2, 0.3, 0.4, 0.5, 0.6, 0.7, 0.8, 0.9, 0.99), and SNR levels (2 dB, 5 dB, 10 dB, 20 dB, ∞). (a) The PEC estimates in the delta band. (b) The relative errors between PEC estimates and simulated ground-truth coupling coefficients. The white line, red line and black line represent the relative error at 10%, 5% and 1%, respectively.

### D. Simulation 2: Discriminative Capacity Assessment

This simulation aimed to test the capacity of PEC in distinguishing true synchronization from random spurious coupling. For each $\lambda$ , we generated 1000 positive samples (synchronized edge pairs constructed as described) and 1000 negative samples (non-synchronized edge pairs created by randomly shuffling the reference-target correspondence). We then computed PEC values for all edge pairs and performed receiver operating characteristic (ROC) analysis to evaluate discriminative performance. Specifically, we defined a threshold $T(p)$ as the $p$-th percentile ($p \in [0, 100], \Delta p = 1$) of the negative-sample PEC distribution [30]. Edge pairs with PEC values exceeding $T(p)$ were identified as truly synchronized. For each threshold, we calculated the true positive rate (TPR) and false positive rate (FPR $= 1 - p$). The area under the ROC curve (AUROC) served as the primary performance metric, with AUROC > 0.7 indicating good discriminative capacity [31] (Fig. 3a).

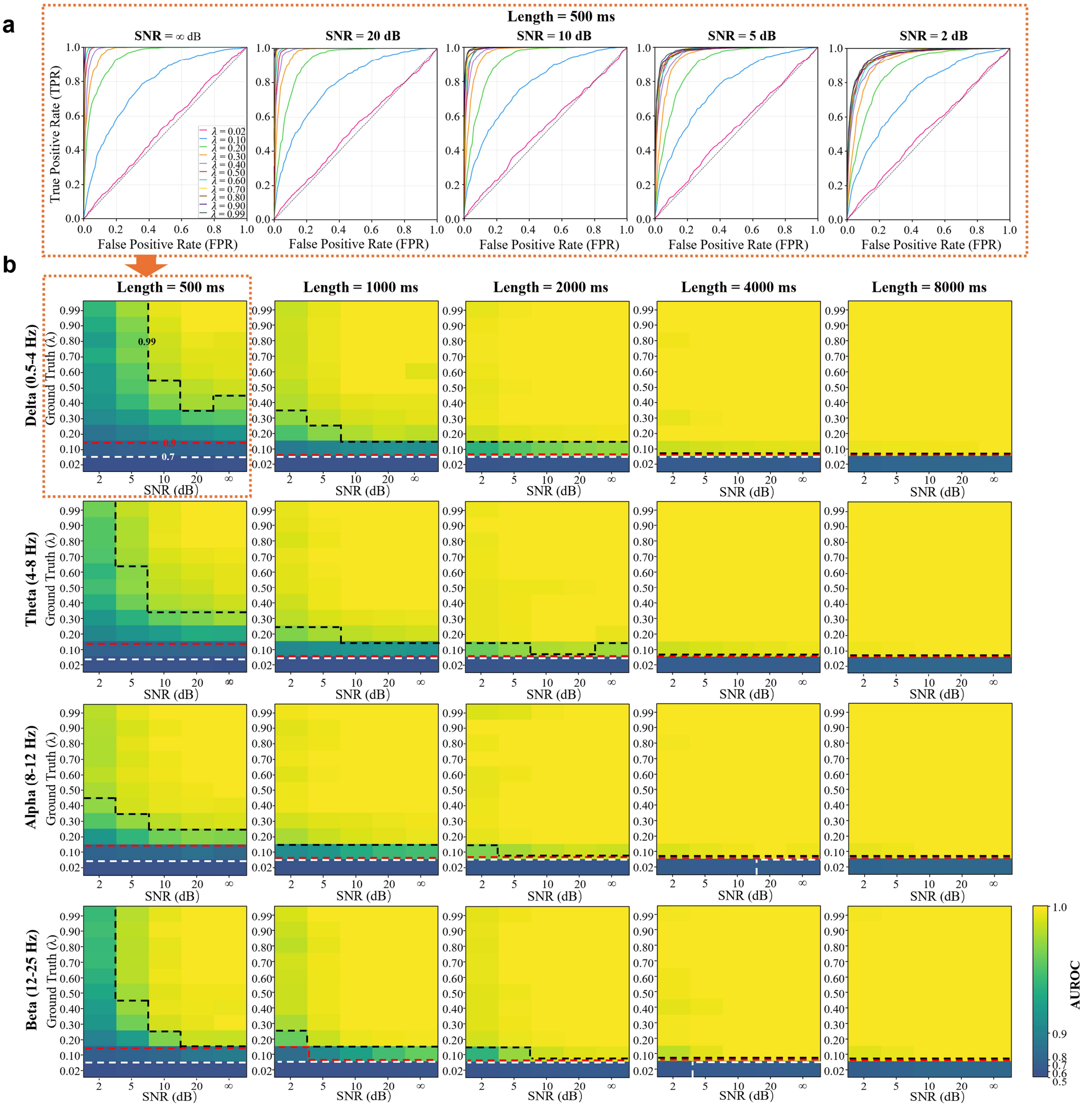

Fig. 3. AUROC values of the PEC algorithm in discriminating true edge coupling from randomly generated spurious coupling. (a) ROC curves in the delta band under a signal length at 500 ms. (b) AUROC values across delta, theta, alpha, and beta bands under varying signal lengths (500 ms, 1000 ms, 2000 ms, 4000 ms, 8000 ms), ground-truth coupling coefficients ($\lambda$ = 0.02, 0.1, 0.2, 0.3, 0.4, 0.5, 0.6, 0.7, 0.8, 0.9, 0.99), and SNR levels (2 dB, 5 dB, 10 dB, 20 dB, ∞). The white line, red line and black line represent AUROC value at 0.7, 0.9 and 0.99, respectively.

It is demonstrated that PEC algorithm also exhibited high reliability in discriminating true edge coupling from randomly generated spurious coupling (Fig. 3b). Overall, the AUROC values exceeded 0.7 in all cases except for the extremely weak coupling condition of $\lambda = 0.02$, with performance improving as coupling coefficient, signal length, and SNR increased. Specifically, at a signal length of 500 ms, the AUROC remained >0.7 for $\lambda \geq 0.1$, and further achieved >0.9 for $\lambda \geq 0.2$. For the signal length of 1000, 2000, and 4000 ms, the AUROC exceeded 0.9 in most cases when $\lambda \geq 0.1$. At the 8000 ms signal length, the AUROC reached >0.7 even under the extremely weak coupling condition $\lambda = 0.02$ and remained >0.9 when $\lambda \geq 0.1$.

With the two simulations, we can conclude that PEC achieved high accuracy in estimating the strength of edge coupling and high reliability in discriminating true edge coupling from randomly generated spurious coupling, demonstrating its robustness in capturing phase synchronization between edges within EEG-derived networks.

## IV. Differential Identifiability

### A. EEG Dataset for Differential Identifiability

To validate the advantage of PEC in brain characterization over traditional PNC, we compared their performances in individual identifiability and cognitive-state identifiability. The dataset used for identifiability assessment (**Dataset 2**) includes dual EEG data from 51 mother-child dyads engaged in a collaborative tangram puzzle task. The dual EEG signals were recorded using two 32-channel NeuSen wireless EEG systems (sampling rate: 1000 Hz, reference: FCz, ground: AFz, impedance <10 kΩ). Specifically, maternal EEG data were selected for identifiability assessment, employing the same preprocessing pipeline as previously described. The dataset was collected in our previous studies [32, 33] approved by the Review Board of Ethics Committee of Shanghai Children's Medical center (SCMCIRB-K2020108-1 and SCMCIRB-K2023099-1).

### B. Individual Identifiability

The differential identifiability metric (denoted as $I_{diff}$) was originally introduced by Amico and Goñi (2018) to compare the effectiveness of different functional connectivity representations in capturing individual-level neural characteristics [34]. This metric has since gained wide acceptance and is now considered a benchmark standard for evaluating the identifiability capacity of functional connectivity methods [19]. In this study, this metric was not only applied to quantify individual identifiability, but also extended to evaluate cognitive-state identifiability, providing a comprehensive assessment of capacity to represent stable individual characteristics and sensitivity to dynamic cognitive state transitions.

For individual identifiability, we included 41 mothers (female, average age = 39.27 ± 7.69 years) with ≥9 artifact-free EEG epochs (2-s duration) during maternal puzzle-moving trials, and computed both PEC and PNC matrices for each trial. Specifically, the alpha band, which is crucial for adults in cognitive functioning, anticipatory and sustained attention [35],was then chosen for brain network construction. Then, we calculated pairwise Pearson correlation coefficients across all possible matrix pairs, categorizing them into: (1) within-subject pairs (from same participant) and (2) between-subject pairs (from different participants).

The within-subject identifiability score ($I_{self}$) was defined as the mean Pearson correlation coefficient across all within-subject matrix pairs, where higher values reflect greater intra-individual stability in connectivity patterns. Conversely, the between-subject identifiability score ($I_{othe}$ ) was defined as the mean Pearson correlation coefficient across all between-subject matrix pairs, with lower values indicate reduced similarity between individuals.

The individual differential identifiability $I_{diff-ind}$ was then calculated as:

$$I_{diff-in} = (I_{self} - I_{othe}) \times 100. \quad (13)$$

A high $I_{diff-in}$ value indicates robust individual-specific neural fingerprinting, as evidenced by high within-subject similarity along with low between-subject similarity.

To establish the statistical significance of $I_{diff-ind}$ differences between PNC and PEC, we performed bootstrap resampling (with replacement) to generate 1,000 subsamples, each containing 12 randomly selected participants (30% of 41). Then we computed $I_{diff-ind}$ for both algorithms per subsample, and conducted paired t-tests.

Before the analysis of differential identifiability, spectral analysis was performed to confirm the role of alpha band in the collaborative tangram puzzle task. The result showed that depressed alpha power during maternal acting periods compared to their observing periods (Fig. 4a), justified as the primary frequency band for differential identifiability analysis.

At the group level, PEC outperformed PNC in individual identifiability, achieving a higher differential identifiability score overall ($I_{diff-ind} = 14.97$ for PEC vs. $I_{diff-ind} = 13.68$ for PNC, n = 41, signal length = 2000 ms, Fig. 4b). This superiority was statistically validated via paired t-tests on bootstrap-resampled subsamples (Cohen's d = 0.65, p < 0.001, signal length = 2000 ms, Fig. 4c).

To investigate the impact of varying signal lengths on the individual identifiability of the two algorithms, we segmented the EEG signals into time windows of varying lengths, ranging from 200 ms to 8000 ms with 200 ms intervals. EEG signals were concatenated when native trial durations were insufficient to generate segments beyond 2000 ms. For each window length, we applied the described bootstrap resampling method (30% sampling rate with replacement and 1000 repetitions) to calculate the distribution of $I_{diff-ind}$ values for both the PEC and PNC algorithms, and performed paired t-tests to assess their statistical significance. We found that PEC outperformed PNC significantly for signal lengths ≥ 1600 ms (Cohen's d > 0.34, p < 0.001, Fig. 4d), with its superiority amplifying as signal length increased. Conversely, this advantage of PEC was absent at short signal lengths, and even showed a significantly lower individual identifiability than PNC at a signal length ≤ 1200 ms (Cohen's d > 0.24, p < 0.001, Fig. 4d).

To reduce the potential influence of noise, we applied principal component analysis (PCA) and reconstructed connectivity matrices using top-ranked principal components to test the performance of differential identifiability. The core idea of PCA is to project the original high-dimensional data onto a set of orthogonal bases—principal components (PCs)—ranked in descending order by the amount of variance they explain [36]. In the context of brain network connectivity, PCs with high explained variance are hypothesized to capture stable, macro-level network architecture or individual-specific neural fingerprints, whereas low-variance PCs likely reflect random noise, artifacts, or measurement errors [34]. As a result, reconstructing the connectivity matrices using only the top-ranked PCs would effectively denoise data while preserving key individualized neural signatures, thereby improving the quality of functional connectivity representation. Specifically, we conducted PCA on the concatenated functional connectivity matrices to obtain variance-ordered PCs, then iteratively reconstructed denoised matrices by incrementally incorporating PCs (from PC$_2$ to PC$_N$, where $N$ denotes the total number of trials) and compared the identifiability metrics for PEC and PNC algorithms at each reconstruction level.

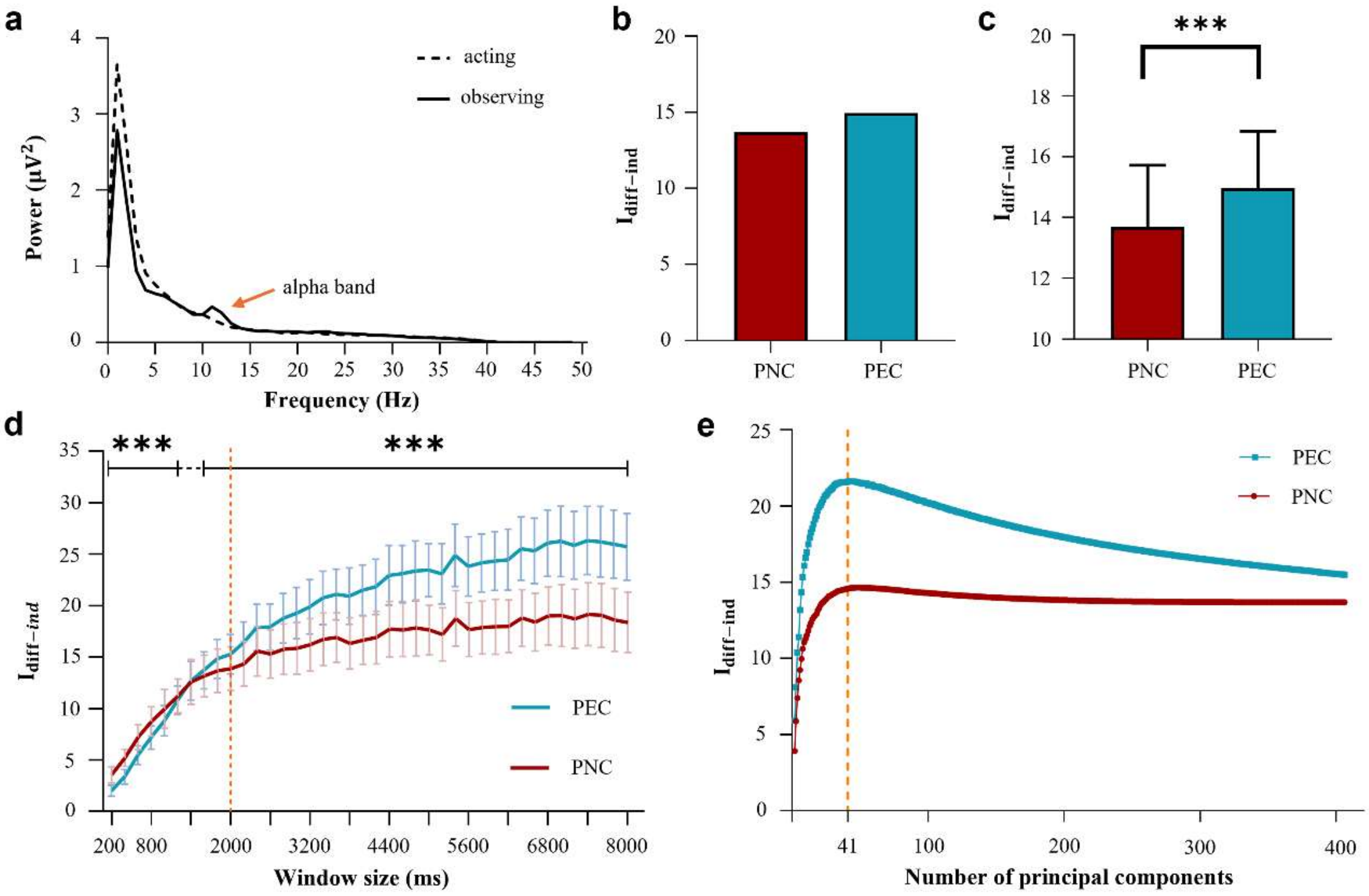

Fig. 4. Comparison of individual identifiability between PEC and PNC. (a) Spectral analysis of maternal EEG during acting and observing. (b) $I_{diff-i}$ for PEC and PNC at the group level. (c) $I_{diff-ind}$ for PEC and PNC on bootstrap-resampled subsamples. (d) $I_{diff-ind}$ for PEC and PNC across segment lengths ranging from 200 to 8000 ms. Error bars represent SD across subsamples. (e) $I_{diff-ind}$ for reconstructed PEC and PNC using different number of principal components. ***p < 0.001.

We found that individual identifiability could be improved by PCA for both PEC and PNC. Both algorithms showed a peak $I_{diff-ind}$ at 41 PCs ($I_{diff-ind} = 21.60$ for PEC vs. $I_{diff-ind} = 14.63$ for PNC), matching our sample size (Fig. 4e). This convergence indicates that the top-ranked PCs encode participant-specific neural signatures. Crucially, PEC maintained a consistent advantage over PNC across all reconstruction levels, demonstrating its robust superiority under optimally denoised conditions.

Finally, we employed 1-Nearest Neighbor (1-NN) classification [37] to evaluate the individual identifiability accuracy, at the best level after PCA reconstruction (41 PCs for both PEC and PNC). Specifically, for each trial (test trial), the algorithm selected the most similar trial (highest Pearson correlation) from the training set (the remaining trials) as its predicted identity. The final accuracy was calculated as the percentage of correctly predicted trials. 1-NN classification results showed a higher individual identification accuracy of PEC (83.30%) than PNC (78.35%), with both algorithms substantially exceeding chance level (1/41 = 2.44%).

## C. Cognitive-state Identifiability

For cognitive-state identifiability assessment, we adapted the differential identifiability framework to evaluate the algorithms' capacity to distinguish cognitive conditions. Specifically, we tried to discriminate two primary experimental states in our turn-taking cooperative puzzle-solving paradigm: the *acting* state (during puzzle-moving) and the *observation* state (during partner's turn). In this analysis, we included 34 mothers who had ≥9 artifact-free EEG epochs (2-s duration) per state.

As in the individual identifiability task, we generated trial-specific PEC/PNC matrices and computed pairwise Pearson correlations, but now categorized them into (1) within-state pairs (from same experimental condition), and (2) between-state pairs (from different experimental conditions). Similarly, the within-state identifiability score ($I_{within}$) represents the mean correlation coefficient across all within-state pairs, and the between-state identifiability score ($I_{between}$) represents the mean correlation coefficient across all between-state pairs. The state differential identifiability $I_{diff-state}$ was then defined as:

$$I_{diff} = (I_{within} - I_{between}) \times 100. \quad (14)$$

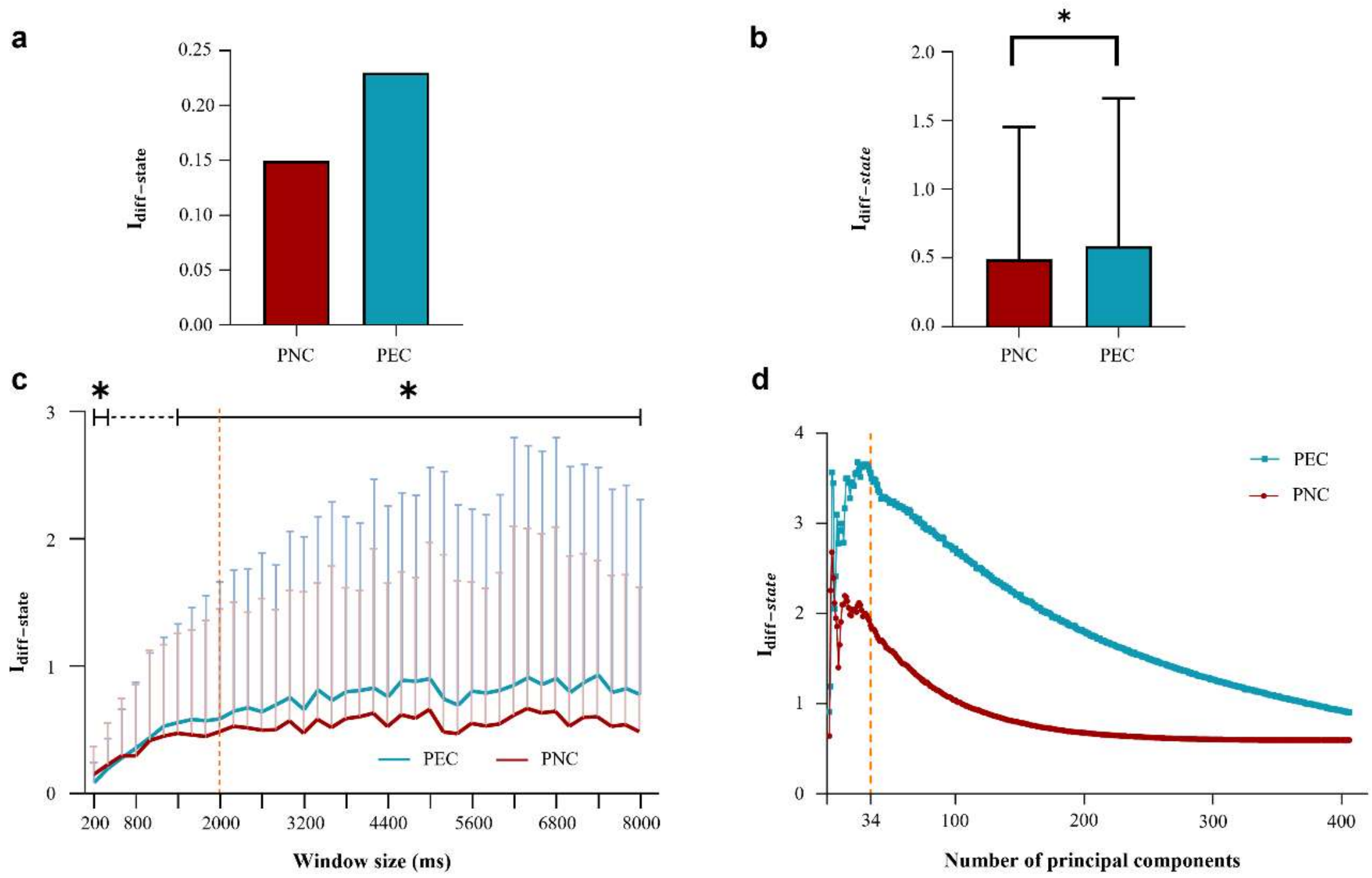

Fig. 5. Comparison of cognitive-state identifiability between PEC and PNC. (a) Cross-subject $I_{diff-stat}$ for PEC and PNC. (b) Within-subject $I_{diff-}$ for PEC and PNC. (c) Within-subject $I_{diff-state}$ for PEC and PNC across segment lengths ranging from 200 to 8000 ms. Error bars represent SD across subjects. (d) Averaged within-subject $I_{diff-stat}$ for reconstructed PEC and PNC from different number of principal components. *p < 0.05.

This metric quantifies an algorithm's ability to represent distinct cognitive states through differentiated connectivity patterns. $I_{diff-}$ was calculated both across subjects and within each subject, and paired t-test was performed on within-subject $I_{diff-state}$ to establish the statistical significance between PNC and PEC performance.

PEC also outperformed PNC in cognitive-state identifiability. At the group level, PEC achieved a higher cross-subject $I_{diff-state}$ ($I_{diff-state} = 0.23$ for PEC vs. $I_{diff-state} = 0.15$ for PNC, n = 34, signal length = 2000 ms, Fig. 5a). To control the confounding factor of individual variability, we calculated the within-subject $I_{diff-state}$ for each subject. Paired t-test demonstrated that PEC showed significantly higher within-subject $I_{diff-sta}$ than PNC (Cohen's d = 0.14, p = 0.04, signal length = 2000 ms; Fig. 5b). Similarly, this superiority of PEC maintained for signal lengths ≥1600 ms (Cohen's d > 0.10, p < 0.05, Fig. 5c), but absent at short signal lengths and even a significantly opposite trend at a signal length = 200 ms (Cohen's d = 0.34, p < 0.05).

Using PCA-based reconstruction, we found that cognitive-state identifiability could also be improved for both PEC and PNC (Fig. 5d). $I_{diff-state}$ of both algorithms presented severe fluctuations at initial PCs, followed by a progressive decline, with peak of PEC around 32-34 PCs ($I_{diff-state}$ = 3.60)—near the number of participants in this state-specific analysis (n = 34)—while peak of PNC 14 PCs ($I_{diff-sta}$ = 2.20). This result further supports the interpretation that high-variance PCs capture subject-specific information. Throughout the reconstruction process, PEC maintained a performance advantage over PNC, which indicated the robustness and reliability of the PEC algorithm in cognitive state recognition performance.

In the 1-NN classification task after PCA reconstruction (34 PCs for PEC and 14 PCs for PNC), PEC achieved a higher accuracy than PNC (85.27% vs. 78.47%) and both performances were greater than the chance level (50%).

In summary, PEC consistently outperformed PNC in both individual identifiability and cognitive-state identifiability under most experimental conditions. Crucially, this performance advantage emerged exclusively under sufficient data duration (≥1600 ms for alpha band), with its magnitude increasing with signal length. These findings underscore fundamentally distinct data-length dependencies between the two algorithms. PNC estimates the average phase synchronization between vertices. This averaging operation provides temporal smoothing for stable phase synchronization estimates even with short data lengths. Previous surrogate test suggested a minimum of 3 oscillatory cycles for PNC estimation [38]. However, PEC constructs edge time series through pointwise phase differentiation, where insufficient signal length (<16 cycles from our results) produces truncated edge sequences, yielding high-variance edge-correlation estimates. These findings align with the fMRI study, suggesting edge-centric models, due to their higher model complexity and information dimensionality, impose stricter requirements on the quantity of input data [19].

By dimensionality reduction and reconstruction after PCA, we found significant improvements in both individual identifiability and cognitive-state identifiability for both PEC and PNC, particularly when retaining PCs equal to the number of subjects in the dataset. The results align with the previous fMRI study [19], confirming that PCA effectively reduced noise while preserving key neural signatures. Notably, PEC consistently outperformed PNC across all tested PCs counts, suggesting that PEC inherently captures more individual and state-dependent discriminative neurophysiological feature, rather from differential denoising effects [34].

## V. Intra-brain Network Community Detection

### A. Leiden Community Detection Algorithm

To evaluate the efficacy of PEC in identifying overlapping communities, we performed intra-brain network community detection analysis using the Leiden algorithm [39]. Among various community detection algorithms, those based on modularity maximization have gained wide adoption in large-scale brain network analysis due to their computational efficiency and performance in handling weighted networks [40]. Especially, the Leiden algorithm provides an advanced variant that address the limitation of poorly connected communities in the standard Louvain method [39], ensuring well-connected and structurally cohesive communities.

The EEG epochs (2-s duration, alpha band) during maternal puzzle-moving trials in **Dataset 2** were used for community detection. To ensure robust network estimation, only 41 mothers with ≥9 artifact-free epochs were included. Before community detection, we computed both PNC and PEC matrices for all the trials, and then these individual trial-level matrices were averaged to generate group-level PNC and PEC matrices. To eliminate spurious connections arising from noise or volume conduction effects while preserving the topologically significant backbone of the network, we applied proportional thresholding [41] by retaining the strongest 20% of connections in the matrices. This threshold was selected within the established optimal range (10%-30%) for balancing biological plausibility and topological stability of brain networks [42].

The sparsified connectivity matrices were then subjected to community detection using the Leiden algorithm. For the PNC matrix, this process directly produces vertex communities; whereas for the PEC matrix, the algorithm first identifies edge communities, which are subsequently mapped to vertex communities by assigning each vertex to all communities containing its connected edges to gather the overlapping structure [4].

Despite the inherent limitation of EEG's low spatial resolution, the sensor-level functional communities inferred from EEG networks were compared against the Yeo 7-network cortical parcellation scheme [43], a template derived from resting-state fMRI data of 1,000 healthy adults, to establish their neuroanatomical validity. This reference framework divides the cerebral cortex into seven canonical functional networks: (1) Visual Network, (2) Somatomotor Network, (3) Dorsal Attention Network, (4) Ventral Attention Network, (5) Limbic Network, (6) Frontoparietal Control Network, and (7) Default Mode Network. The Yeo parcellation has demonstrated exceptional robustness across diverse populations and experimental conditions, recognized as a gold standard for functional brain organization [44].

### B. Comparison of PEC and PNC in Intra-brain Network Community Detection

PNC network during maternal puzzle-moving yielded five non-overlapping vertex communities, while PEC matrix identified four edge communities, from which four corresponding overlapping vertex communities were derived through edge-to-vertex projection (Fig. 6). When compared against the canonical Yeo 7-network parcellation [43], PEC modelling demonstrated sensor-level topographic correspondence with established functional systems.

**Lateral Fronto-Centro-Parietal Community (Fig. 6a).** PEC identified a unified community (PEC-C1) spanning lateral frontal, central and parietal electrodes (PEC-C1: F7, F8, FC5, FC6, C3, C4, T7, T8, CP5, CP6, P8), which spatially overlapped with the integration of the Somatomotor Network (SMN) and Frontoparietal Control Network (FPCN). This integrated architecture aligns with neurocognitive models positing that goal-directed motor planning requires tight coupling between FPCN-mediated cognitive control and SMN-mediated motor execution [45, 46]. Critically, PNC failed to capture this functional unity, instead partitioning these regions into two hemispherically segregated communities (PNC-C1 for left hemisphere, PNC-C2 for right hemisphere), thereby obscuring the bilateral integration essential for coordinated goal-directed motor planning during tangram manipulation.

**Medial-Anterior Frontal Community (Fig. 6b).** PEC Network. This spatial pattern may reflect the limbic system's[47], which are particularly salient in the mother-child collaborative paradigm where social bonding and task-related reward (e.g., puzzle completion success) likely engage this network [48]. In contrast, PNC fragmented this system into two isolated components (PNC-C3: Fp1, Fp2; PNC-C4: Fz), thereby disrupting the functional coherence of this anterior frontal module.

**Central-Parieto-Occipital Community (Fig. 6c).** PEC identified a central-parieto-occipital community (PEC-C3: C3, C4, Cz, CP1, CP2, CP5, CP6, P3, P4, P7, P8, Pz, O1, O2, Oz) that spatially aligned with the integration of SMN and Visual Network (VN). This module likely reflects the dorsal visual stream's role in transforming visual input into motor commands [49], which is critical for visually guided puzzle manipulation (e.g., shape recognition, spatial rotation, and hand-eye coordination). While PNC partially recovered this posterior cluster (PNC-C5), it excluded key electrodes (P7, P8, CP5, CP6) that had been allocated to its first two communities (PNC-C1/C2), resulting in an anatomically incomplete representation of this sensorimotor-visual integration system.

**Bilateral Frontal Community (Fig. 6d).** PEC uniquely identified a community (PEC-C4: F3, F4, F7, F8, FC1, FC2, FC5, FC6) that showed topological correspondence with the canonical FPCN. This community likely supports dual-task demands during the tangram task, such as maintaining the target puzzle configuration in working memory while simultaneously monitoring children's actions and emotional state—a hallmark of FPCN-mediated adaptive control [50, 51]. PNC failed to detect this module, as its vertices were exclusively assigned to the somatomotor-integration communities (PNC-C1/C2), illustrating a fundamental limitation of non-overlapping partitioning in capturing context-dependent network reconfiguration.

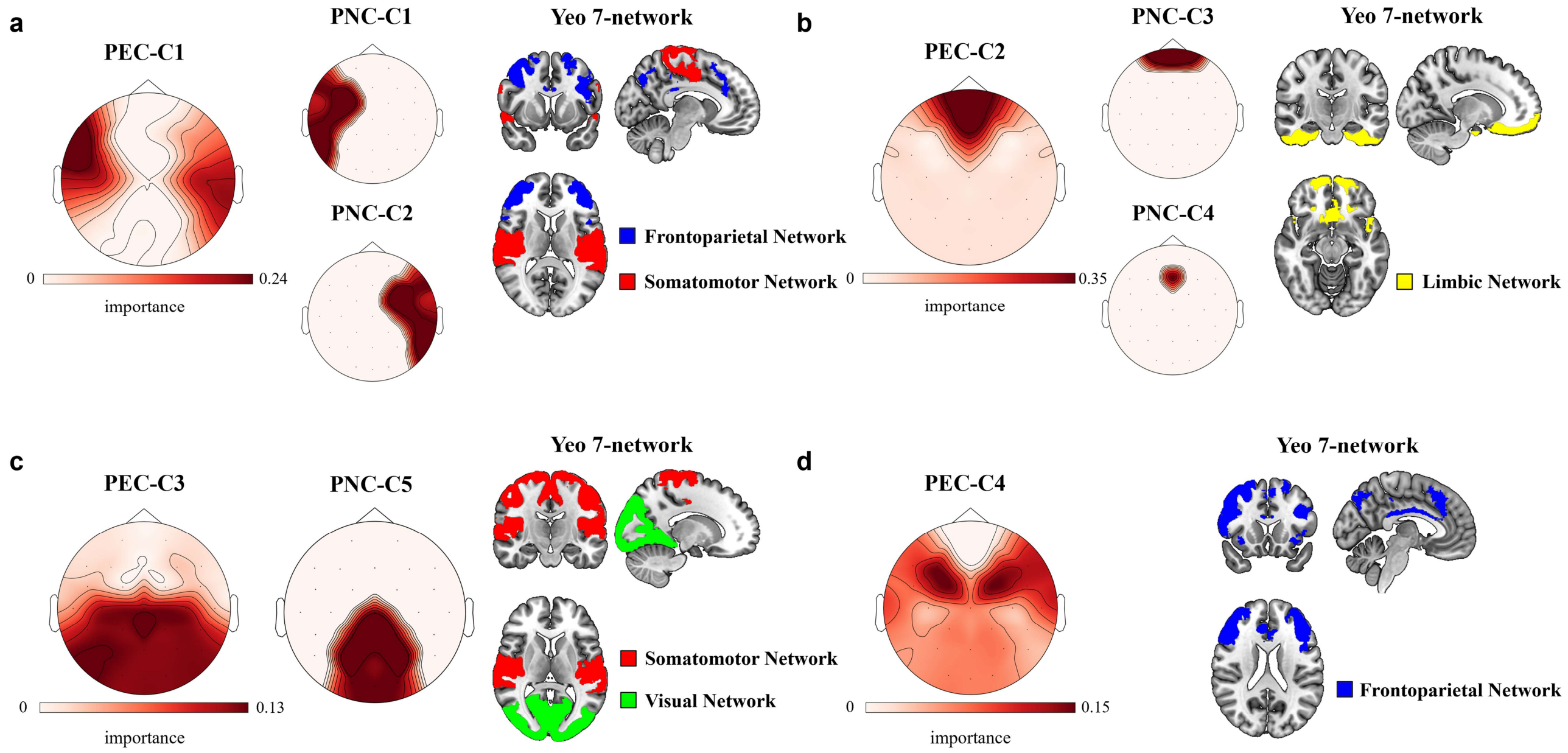

Fig. 6. Four communities found in maternal PEC network during mother-child cooperation. (a) Lateral Fronto-Centro-Parietal Community. (b) Medial-Anterior Frontal Community. (c) Central-Parieto-Occipital Community. (d) Bilateral Frontal Community. In each subfigure, the left column stands for the detected community in the PEC network, where the colorbar indicates the importance of nodes to the community (node importance is quantified as the ratio of edges connecting the node to all edges within the community); the middle column stands for the detected community/communities in the PNC network; the right column stands for the corresponding Yeo 7-network parcellation.

To conclude, the PEC-based network revealed a highly overlapping yet functionally specialized modular architecture, which aligns the cognitive processes engaged during the task.However, the PNC-based analysis, which enforces a non-overlapping partition where each node belongs to a single community, failed to capture this nuanced architecture. Critically, PEC allows us to identify connector hubs that dynamically bridge multiple functional systems. For example, in our tangram solving task, the SMN (including electrodes C3, C4, CP5, CP6, P8) served as a shared component between Module 1 and Module 3, while the FPCN (including F7, F8, FC5, FC6) overlapped between Module 1 and Module 4. These findings align with established network neuroscience principles, where the FPCN acts as a flexible hub for dynamic cognitive control, and the SMN serves as the core executor in the perception-action loop [43, 51]. It is important to note that while we utilized the Yeo atlas for functional annotation based on spatial topography, the sensor-level resolution inherent to EEG necessitates interpreting these mappings as of scalp projection rather than precise cortical sources. Nevertheless, the similarity between our source-level communities and the canonical fMRI networks suggests that PEC probably captures meaningful large-scale functional architectures as well as reveals the brain's overlapping, and multifunctional organization.

## VI. Inter-brain Network Community Detection

### A. Bipartite Network Community Detection Using the BRIM Algorithm

To examine edge-level mother-child neural coupling, we constructed a dyadic PEC matrix for each trial by computing PEC values between all possible mother-child edge pairs. Then, the group-level PEC matrix was obtained by averaging across trials and dyads, followed by 20% sparsity thresholding to retain robust inter-brain connections. Given the bipartite structure of inter-brain edge networks (edges from two brains form non-overlapping node sets), we applied the BRIM algorithm [52] to the group-level dyadic PEC matrix to identify cross-brain modules. This algorithm optimizes bipartite modularity through iterative alternating optimization, partitioning the two distinct node sets (child-edges and mother-edges) into communities that maximize intra-module connectivity while minimizing inter-module connectivity.

### B. Modular Architecture of Mother-child Inter-brain Edge Network

The BRIM algorithm yielded independent community assignments for each brain, identifying three edge communities in child's network and three in mother's network, which collectively formed nine cross-brain modules (Fig. 7a). Of these, five modules showed null inter-brain connectivity (all PEC values = 0), and the remaining four non-zero modules were designated as Module 1-4. Specifically, Module 1 was characterized by its strongest, fully-connected inter-brain connectivity, with all its PEC values exceeding 0.4 and a density peak in the range of 0.4–0.7. Modules 2 and 3 were featured by its sparse strong connectivity, with their PEC distributions exhibiting a bimodal profile: a high-synchrony cluster (0.4–0.5) and a low-synchrony cluster (0–0.1). Module 4 was composed of uniformly weak connections (0–0.03) (Fig. 7b).

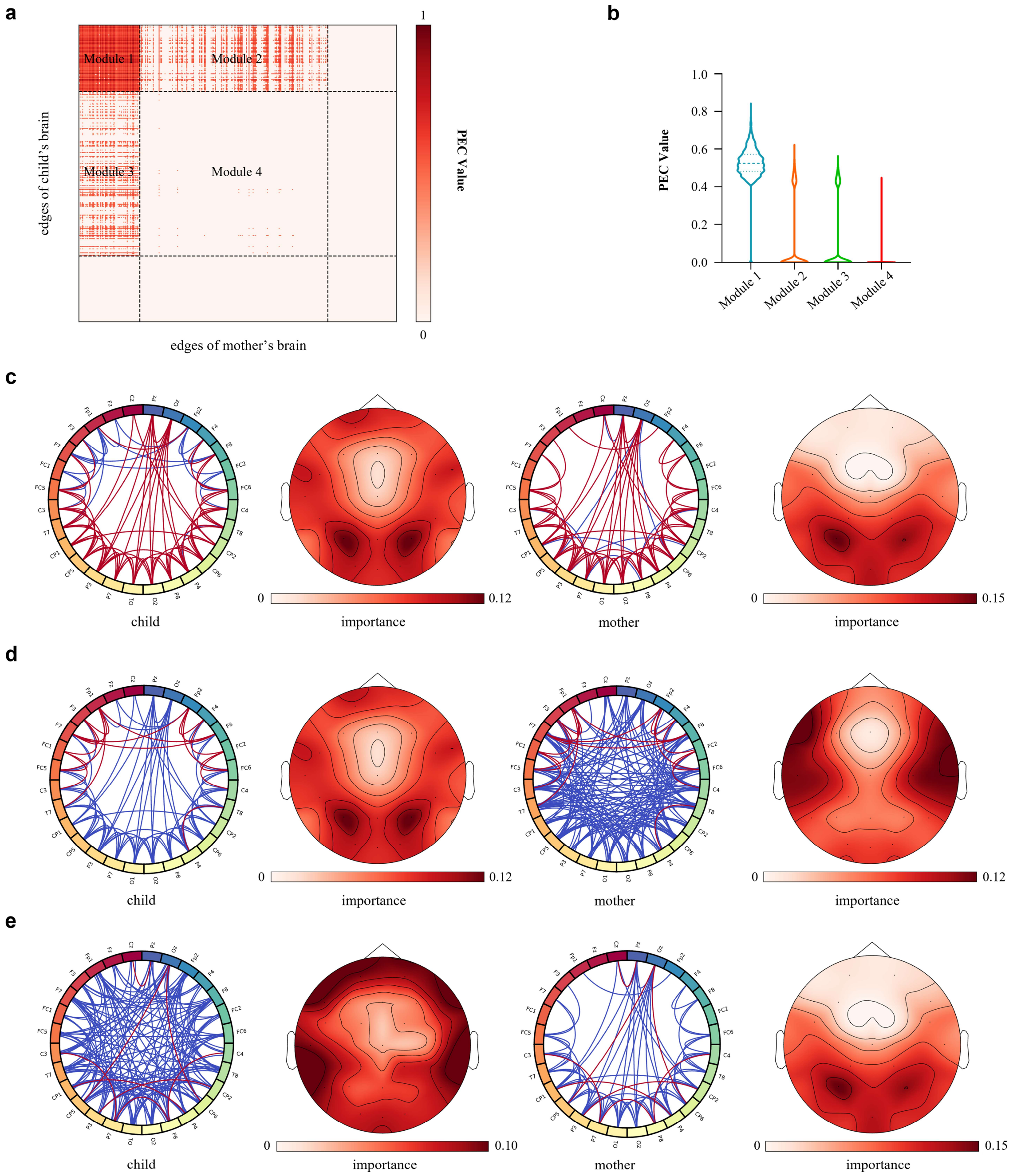

Fig. 7. Modular architecture of inter-brain edge network during mother-child cooperation. (a) Cross-brain community detection using the BRIM algorithm. (b) PEC distributions of inter-brain edge modules. (c-e) Brain connections and their topological distributions in Module 1 (c), Module 2 (d), and Module 3 (e). In the network visualizations, the red lines represent shared edges in mother's and child's network, while the blue lines represent specific connections unique to each individual. The colorbars indicate the importance of nodes to the community, which is quantified as the ratio of edges connecting the node to all edges within the community.

To further elucidate the neural substrates of these modules, we visualized the dominant brain edges within Modules 1–3. In Module 1, the edges from the child and mother demonstrated a high topological similarity (Intersection-over-Union [IoU] score: 0.70), reinforcing the module's symmetric nature (Fig. 7c). These shared edges were predominantly localized in posterior brain regions, with parietal (P3/P4) and occipital (O1/O2) electrodes acting as central hubs. This symmetric co-activation module likely reflects shared visual processing and mirrored neural responses between actor and observer during cooperative puzzle solving. Dyadic partners received identical visual stimuli during the task, driving coordinated activations in the occipital cortex. Crucially, while the actor engaged sensorimotor networks for physical manipulation, the observer recruited analogous neural circuits through "motor imagery"—a neural mechanism where action observation activates similar network [53].

In contrast, Modules 2 and 3 demonstrated asymmetric connectivity architectures, with little overlap between the child's and mother's edges (both IoU < 0.09). Module 2 exhibited child's posterior – mother's anterior coupling, i.e., the child's edges primarily in the parietal (P3/P4) and occipital (O1/O2) electrodes coupled with the mother's edges concentrated in frontal (F7/F8, FC5/FC6) and temporal (T7/T8) areas (Fig. 7d). Similarly, Module 3 exhibited child's anterior – mother's posterior coupling, where the child's edges in the frontal (Fp1/Fp2, F7/F8, FC5/FC6) and temporal (T7/T8, CP5/CP6) electrodes coupled with the mother's edges in the parietal (P3/P4) and occipital (O1/O2) electrodes (Fig. 7e). These two asymmetric modules like encode the distinct neurocognitive roles of mother and child (direct action execution versus observation) during the cooperative task. Although actor and observer may share similar activations in some part of network (Module 1), their distinct roles in interaction may activate different regions. It has been proposed that two interacting brains constitute a complex system wherein inter-brain synchrony occurs not only between homologous regions but also among non-homologous areas [16].

From the results of inter-brain network community detection, we can conclude that PEC effectively decomposed cross-brain modules encoding the symmetric and asymmetric interactive neurocognitive processes during dyadic interaction. This computational framework advances our understanding of how mother-child neural systems coordinate to support cooperative social engagement.

## VII. Advantages and Limitations

In this study, we developed a phase-based edge coupling algorithm by constructing edge time series from instantaneous phase differences of EEG signals and quantifying edge-to-edge coupling via normalized inner product, addressing the methodological gap in edge-centric connectivity modeling for EEG network analysis. Simulation tests confirmed robustness of our algorithm, generally achieving <10% relative error in edge coupling strength estimation across frequency bands and AUROC >0.7 in discriminating synchronized/non-synchronized edge pairs, except at the extremely weak coupling condition ($\lambda = 0.02$). PEC significantly outperformed traditional node-centric PNC approaches in individual identifiability and cognitive-state identifiability, though these advantages required sufficient data length (>16 cycles) due to its edge-centric complexity. Unlike PNC's hard-partitioned modularity, PEC showed potential to capture highly overlapping yet functionally specialized modular structures. It successfully identified four key modules that align with tangram-solving cognitive processes, including connector hubs (e.g., somatomotor and frontoparietal networks) that bridge functional systems. When applied to inter-brain analysis, PEC effectively decomposed cross-brain modules encoding the symmetric and asymmetric interactive neurocognitive processes during dyadic interaction, providing novel insights into the neural substrates of cooperative engagement.

Although PEC algorithm demonstrates multiple advantages in brain functional characterization, several limitations must be acknowledged. First, edge networks exhibit inherent high-dimensional complexity compared to node-centric representations, with edge counts scaling quadratically with the number of nodes ($O(N^2)$). This dimensionality explosion generates substantial increases in storage requirements, computational load, and analytical overhead, posing challenges for large-scale neuroimaging studies (e.g., source level) or real-time applications. Second, PEC's computational complexity necessitates stringent data length requirements, as insufficient signal duration leads to truncated edge time series and high-variance coupling estimates. This constraint may limit its application to brief event-related paradigms where edge-centric modeling may fail to capture transient neural dynamics. Third, the current PEC exclusively quantifies phase-based coupling, constructing edges from instantaneous phase differences of nodal signals, and measuring edge-to-edge coupling from temporal consistency of their phase relationships, thereby inherently overlooking alternative neural synchronization mechanisms such as amplitude co-modulation. Future work should integrate diverse coupling metrics into the edge-centric framework.

## Acknowledgment

This work was supported by the National Natural Science Foundation of China (No. 62371285) and Fundamental Research Funds for the Central Universities (No. YG2023QNB21). We thank Siyang Liu and Xiyuan Lei for their help in data preprocessing.

## References

[1] E. Bullmore, and O. Sporns, "Complex brain networks: graph theoretical analysis of structural and functional systems," *Nat Rev Neurosci,* vol. 10, no. 3, pp. 186-98, Mar, 2009.

[2] S. Achard, and E. Bullmore, "Efficiency and cost of economical brain functional networks," *PLoS Comput Biol,* vol. 3, no. 2, pp. e17, Feb 2, 2007.

[3] O. Sporns, "Network attributes for segregation and integration in the human brain," *Curr Opin Neurobiol,* vol. 23, no. 2, pp. 162-71, Apr, 2013.

[4] J. Faskowitz, F. Z. Esfahlani, Y. Jo, O. Sporns, and R. F. Betzel, "Edge-centric functional network representations of human cerebral cortex reveal overlapping system-level architecture," *Nat Neurosci,* vol. 23, no. 12, pp. 1644-1654, Dec, 2020.

[5] M. A. Bertolero, B. T. T. Yeo, D. S. Bassett, and M. D'Esposito, "A mechanistic model of connector hubs, modularity and cognition," *Nat Hum Behav,* vol. 2, no. 10, pp. 765-777, Oct, 2018.

[6] A. V. Utevsky, D. V. Smith, and S. A. Huettel, "Precuneus Is a Functional Core of the Default-Mode Network," *Journal of Neuroscience,* vol. 34, no. 3, pp. 932-940, Jan 15, 2014.

[7] H. Zhu, W. Jin, J. Zhou, S. B. Tong, X. K. Xu, and J. F. Sun, "Nodal Memberships to Communities of Functional Brain Networks Reveal Functional Flexibility and Individualized Connectome," *Cerebral Cortex,* vol. 31, no. 11, pp. 5090-5106, Nov, 2021.

[8] G. Palla, I. Derenyi, I. Farkas, and T. Vicsek, "Uncovering the overlapping community structure of complex networks in nature and society," *Nature,* vol. 435, no. 7043, pp. 814-8, Jun 9, 2005.

[9] P. K. Gopalan, and D. M. Blei, "Efficient discovery of overlapping communities in massive networks," *Proc Natl Acad Sci U S A,* vol. 110, no. 36, pp. 14534-9, Sep 3, 2013.

[10] J. Yang, and J. Leskovec, "Overlapping Communities Explain Core–Periphery Organization of Networks," *Proceedings of the IEEE,* vol. 102, no. 12, pp. 1892-1902, 2014.
[11] L. Hao, "Overlapping community detection with least replicas in complex networks," *Information Sciences,* vol. 453, pp. 216-226, 2018.
[12] J. Faskowitz, R. F. Betzel, and O. Sporns, "Edges in brain networks: Contributions to models of structure and function," *Netw Neurosci,* vol. 6, no. 1, pp. 1-28, Feb, 2022.
[13] S. E. Morgan, S. Achard, M. Termenon, E. T. Bullmore, and P. E. Vertes, "Low-dimensional morphospace of topological motifs in human fMRI brain networks," *Netw Neurosci,* vol. 2, no. 2, pp. 285-302, 2018.
[14] J. H. Balsters, V. Zerbi, J. Sallet, N. Wenderoth, and R. B. Mars, "Primate homologs of mouse cortico-striatal circuits," *Elife,* vol. 9, Apr 16, 2020.
[15] V. Muller, "Neural Synchrony and Network Dynamics in Social Interaction: A Hyper-Brain Cell Assembly Hypothesis," *Front Hum Neurosci,* vol. 16, pp. 848026, 2022.
[16] C. Gerloff, K. Konrad, D. Bzdok, C. Busing, and V. Reindl, "Interacting brains revisited: A cross-brain network neuroscience perspective," *Hum Brain Mapp,* vol. 43, no. 14, pp. 4458-4474, Oct 1, 2022.
[17] Y. Y. Ahn, J. P. Bagrow, and S. Lehmann, "Link communities reveal multiscale complexity in networks," *Nature,* vol. 466, no. 7307, pp. 761-4, Aug 5, 2010.
[18] M. A. de Reus, V. M. Saenger, R. S. Kahn, and M. P. van den Heuvel, "An edge-centric perspective on the human connectome: link communities in the brain," *Philos Trans R Soc Lond B Biol Sci,* vol. 369, no. 1653, Oct 5, 2014.
[19] Y. Jo, J. Faskowitz, F. Z. Esfahlani, O. Sporns, and R. F. Betzel, "Subject identification using edge-centric functional connectivity," *Neuroimage,* vol. 238, Sep, 2021.
[20] L. Novelli, and A. Razi, "A mathematical perspective on edge-centric brain functional connectivity," *Nat Commun,* vol. 13, no. 1, pp. 2693, May 16, 2022.
[21] H. M. Jones, K. Yoo, M. M. Chun, and M. D. Rosenberg, "Edge-Based General Linear Models Capture Moment-to-Moment Fluctuations in Attention," *J Neurosci,* vol. 44, no. 14, Apr 3, 2024.
[22] F. Zamani Esfahlani, L. Byrge, J. Tanner, O. Sporns, D. P. Kennedy, and R. F. Betzel, "Edge-centric analysis of time-varying functional brain networks with applications in autism spectrum disorder," *Neuroimage,* vol. 263, pp. 119591, Nov, 2022.
[23] B. Cai, G. Zhang, W. Hu, A. Zhang, P. Zille, Y. Zhang, J. M. Stephen, T. W. Wilson, V. D. Calhoun, and Y. P. Wang, "Refined measure of functional connectomes for improved identifiability and prediction," *Hum Brain Mapp,* vol. 40, no. 16, pp. 4843-4858, Nov 1, 2019.
[24] J. P. Lachaux, E. Rodriguez, J. Martinerie, and F. J. Varela, "Measuring phase synchrony in brain signals," *Human Brain Mapping,* vol. 8, no. 4, pp. 194-208, 1999.
[25] E. Kreyszig, *Introductory Functional Analysis with Applications*, 1989.
[26] X. Lei, A. Wang, K. Zhang, S. Liu, Y. Zhao, S. Laureys, S. Tong, H. Di, N. Hu, and X. Guo, "Characterizing consciousness states: EEG microstate dynamics in patients with disorders of consciousness during naturalistic movie-viewing," *Med Biol Eng Comput,* vol. 63, no. 12, pp. 3695-3707, Dec, 2025.
[27] A. Delorme, and S. Makeig, "EEGLAB: an open source toolbox for analysis of single-trial EEG dynamics including independent component analysis," *J Neurosci Methods,* vol. 134, no. 1, pp. 9-21, Mar 15, 2004.
[28] K. V. Mardia, and P. E. Jupp, *Directional Statistics*, 1999.
[29] M. Abramowitz, and I. A. Stegun, *Handbook of Mathematical Functions with Formulas, Graphs, and Mathematical Tables*, 1964.
[30] A. Basti, F. Chella, R. Guidotti, M. Ermolova, A. D'Andrea, M. Stenroos, G. L. Romani, V. Pizzella, and L. Marzetti, "Looking through the windows: a study about the dependency of phase-coupling estimates on the data length," *Journal of Neural Engineering,* vol. 19, no. 1, Feb 1, 2022.
[31] J. A. Swets, "Measuring the Accuracy of Diagnostic Systems," *Science,* vol. 240, no. 4857, pp. 1285-1293, Jun 3, 1988.
[32] Y. Li, S. Wu, J. Xu, H. Wang, Q. Zhu, W. Shi, Y. Fang, F. Jiang, S. Tong, Y. Zhang, and X. Guo, "Interbrain substrates of role switching during mother-child interaction," *Hum Brain Mapp,* vol. 45, no. 5, pp. e26672, Apr, 2024.
[33] J. Xu, Y. Li, R. Su, S. Wu, C. Wu, H. Wang, Q. Zhu, Y. Fang, F. Jiang, S. Tong, Y. Zhang, and X. Guo, "Dynamic Theta-Alpha Inter-Brain Model during Mother-Preschooler Cooperation," *IEEE J Biomed Health Inform,* vol. 30, no. 2, pp. 1060-1072, Feb, 2026.
[34] E. Amico, and J. Goñi, "The quest for identifiability in human functional connectomes," *Scientific Reports,* vol. 8, May 29, 2018.
[35] O. M. Bazanova, and D. Vernon, "Interpreting EEG alpha activity," *Neurosci Biobehav Rev,* vol. 44, pp. 94-110, Jul, 2014.
[36] L. Billard, and E. Diday, *Principal Component Analysis*, 2006.
[37] E. S. Finn, X. Shen, D. Scheinost, M. D. Rosenberg, J. Huang, M. M. Chun, X. Papademetris, and R. T. Constable, "Functional connectome fingerprinting: identifying individuals using patterns of brain connectivity," *Nat Neurosci,* vol. 18, no. 11, pp. 1664-71, Nov, 2015.
[38] S. Junfeng, H. Xiangfei, and T. Shanbao, "Phase synchronization analysis of EEG signals: an evaluation based on surrogate tests," *IEEE Trans Biomed Eng,* vol. 59, no. 8, pp. 2254-63, Aug, 2012.
[39] V. A. Traag, L. Waltman, and N. J. van Eck, "From Louvain to Leiden: guaranteeing well-connected communities," *Sci Rep,* vol. 9, no. 1, pp. 5233, Mar 26, 2019.
[40] V. D. Blondel, J.-L. Guillaume, R. Lambiotte, and E. Lefebvre, "Fast unfolding of communities in large networks," *Journal of Statistical Mechanics: Theory and Experiment,* vol. 2008, 2008.
[41] M. Rubinov, and O. Sporns, "Complex network measures of brain connectivity: uses and interpretations," *Neuroimage,* vol. 52, no. 3, pp. 1059-69, Sep, 2010.
[42] K. A. Garrison, D. Scheinost, E. S. Finn, X. Shen, and R. T. Constable, "The (in)stability of functional brain network measures across thresholds," *Neuroimage,* vol. 118, pp. 651-61, Sep, 2015.
[43] B. T. Yeo, F. M. Krienen, J. Sepulcre, M. R. Sabuncu, D. Lashkari, M. Hollinshead, J. L. Roffman, J. W. Smoller, L. Zollei, J. R. Polimeni, B. Fischl, H. Liu, and R. L. Buckner, "The organization of the human cerebral cortex estimated by intrinsic functional connectivity," *J Neurophysiol,* vol. 106, no. 3, pp. 1125-65, Sep, 2011.
[44] S. B. Eickhoff, B. T. T. Yeo, and S. Genon, "Imaging-based parcellations of the human brain," *Nat Rev Neurosci,* vol. 19, no. 11, pp. 672-686, Nov, 2018.
[45] S. Maruyama, M. Fukunaga, S. K. Sugawara, Y. H. Hamano, T. Yamamoto, and N. Sadato, "Cognitive control affects motor learning through local variations in GABA within the primary motor cortex," *Scientific Reports,* vol. 11, no. 1, Sep 17, 2021.
[46] M. Abe, and T. Hanakawa, "Functional coupling underlying motor and cognitive functions of the dorsal premotor cortex," *Behavioural Brain Research,* vol. 198, no. 1, pp. 13-23, Mar 2, 2009.
[47] P. J. Morgane, J. R. Galler, and D. J. Mokler, "A review of systems and networks of the limbic forebrain/limbic midbrain," *Progress in Neurobiology,* vol. 75, no. 2, pp. 143-160, Feb, 2005.
[48] K. C. Berridge, and M. L. Kringelbach, "Pleasure Systems in the Brain," *Neuron,* vol. 86, no. 3, pp. 646-664, May 6, 2015.
[49] M. A. Goodale, and A. D. Milner, "Separate Visual Pathways for Perception and Action," *Trends in Neurosciences,* vol. 15, no. 1, pp. 20-25, Jan, 1992.
[50] E. K. Miller, and J. D. Cohen, "An integrative theory of prefrontal cortex function," *Annual Review of Neuroscience,* vol. 24, pp. 167-202, 2001.
[51] M. W. Cole, J. R. Reynolds, J. D. Power, G. Repovs, A. Anticevic, and T. S. Braver, "Multi-task connectivity reveals flexible hubs for adaptive task control," *Nature Neuroscience,* vol. 16, no. 9, pp. 1348-U247, Sep, 2013.
[52] M. J. Barber, "Modularity and community detection in bipartite networks," *Physical Review E,* vol. 76, no. 6, Dec, 2007.
[53] M. Lotze, P. Montoya, M. Erb, E. Hulsmann, H. Flor, U. Klose, N. Birbaumer, and W. Grodd, "Activation of cortical and cerebellar motor areas during executed and imagined hand movements: an fMRI study," *J Cogn Neurosci,* vol. 11, no. 5, pp. 491-501, Sep, 1999.